\documentclass[onecolumn, 12pt]{IEEEtran}
\usepackage{setspace}
\usepackage{amsthm}

\ifCLASSINFOpdf
   \usepackage[pdftex]{graphicx}
   \graphicspath{{../pdf/}{../jpeg/}}
   \DeclareGraphicsExtensions{.pdf,.jpeg,.png}
\else
   \usepackage[dvips]{graphicx}
   \graphicspath{{../eps/}}
   \DeclareGraphicsExtensions{.eps}
\fi
\usepackage{epstopdf}
\usepackage{amsmath}
\usepackage{bm}

\usepackage{subfigure}
\usepackage{caption}

\usepackage{amssymb}

\begin{document}
\captionsetup[figure]{labelfont={default},labelformat={default},labelsep=period,name={Fig.}}
\title{Analog Versus Hybrid Precoding for Multiuser Massive MIMO with Quantized CSI Feedback}
%
%
\author{Yaqiong Zhao, \emph{Student Member, IEEE},
       Wei Xu, \emph{Senior Member, IEEE},
       Jindan Xu, \emph{Student Member, IEEE},
       Shi Jin, \emph{Senior Member, IEEE},
       Kezhi Wang, \emph{Member, IEEE},
      and Mohamed-Slim Alouini, \emph{Fellow, IEEE}
\thanks{Y. Zhao, J. Xu, and S. Jin are with the National Mobile Communications Research Laboratory, Southeast University, Nanjing 210096, China (email: $\lbrace$zhaoyaqiong, jdxu, jinshi$\rbrace$@seu.edu.cn). }
\thanks{W. Xu is with the National Mobile Communications Research Laboratory, Southeast University, Nanjing 210096, China, and is also with the Purple Mountain Laboratories, Nanjing 210000, China (wxu@seu.edu.cn).}
\thanks{K. Wang is with the Department of Computer and Information Sciences, Northumbria University, Newcastle, UK (email: kezhi.wang@northumbria.ac.uk).}
\thanks{M. -S. Alouini are with the Computer, Electrical and Mathematical  Science and Engineering Division, King Abdullah University of Science and  Technology, Thuwal, Saudi Arabia, 23955 (email: slim.alouini@kaust.edu.sa).}
}

\maketitle
\begin{abstract}
In this letter, we study the performance of a downlink multiuser massive multiple-input multiple-output (MIMO) system with sub-connected structure over limited feedback channels. Tight rate approximations are theoretically analyzed for the system with pure analog precoding and hybrid precoding. The effect of quantized analog and digital precoding is characterized in the derived expressions. Furthermore, it is revealed that the pure analog precoding outperforms the hybrid precoding using maximal-ratio transmission (MRT) or zero forcing (ZF) under certain conditions, and we theoretically characterize the conditions in closed form with respect to signal-to-noise ratio (SNR), the number of users and the number of feedback bits. Numerical results verify the derived conclusions on both Rayleigh
channels and mmWave channels.

\end{abstract}


%
\IEEEpeerreviewmaketitle
\section{Introduction}
\IEEEPARstart{M}{assive} multiple-input multiple-output (MIMO) has attracted increasing attention as a key technology in the fifth-generation (5G) network [1]. With its advantages for inter-user interference cancellation and noise supression, massive MIMO can achieve near-optimal performance with simple linear precoding schemes, such as  maximal-ratio transmission (MRT) and zero forcing (ZF) [2]. These fully-digital precoding schemes need each antenna to be driven by one dedicated radio-frequency (RF) chain, which imposes prohibitively high cost and power consumption [3]. To address this issue, two kinds of designs have been introduced, i.e., pure analog precoding and hybrid analog-and-digital precoding [4] [5].

Although the hybrid precoding has been widely considered in research, evidence has shown that it may not be the best choice for all cases compared with the pure analog precoding, especially when considering that analog precoding consumes less power than the hybrid precoding due to some implementation facts [4] [6]. In [7] [8], the analog precoding approximately transformed the effective channel into a diagonal matrix, which implies that digital processing is no longer needed for further multiuser interference cancellation. Specifically in [9], the authors found that pure analog processing could surpass the hybrid processing with maximal-ratio combination (MRC) or ZF in an uplink channel under the assumption of perfect channel state information (CSI). For the downlink channel, however, this compromise is still unclear. Moreover in practice, only quantized CSI, instead of perfect one, is available through limited feedback [10]-[13].

Against the above background, this letter investigates the performance of a downlink massive MIMO system with low-cost sub-connected architecture and quantized CSI feedback. Tight rate approximations are derived for the ZF/MRT-based hybrid precodings and the analog precoding in the large base station (BS) antenna regime. With the derived results, we explicitly characterize the performance comparison between the hybrid precoding and the analog precoding. In particular, we show that for all SNRs analog precoding outperforms the ZF-based hybrid precoding when $B_2 \leq B_2^0$ where $B_2$ denotes the number of feedback bits in digital precoding and $B_2^0$ is a constant in closed-form with respect to system parameters. Similar observation has also been obtained for the comparison of MRT-based hybrid precoding and pure analog precoding. For other cases, the superiority of different precoding schemes would be complicated and it depends on specific values of SNR, which has also been derived in closed forms.

The remainder of this paper is organized as follows. System model is introduced in Section \uppercase\expandafter{\romannumeral2}. In Section \uppercase\expandafter{\romannumeral3}, we derive the achievable rates of the system using  various precoding schemes. In Section \uppercase\expandafter{\romannumeral4}, we present the conditions under which the pure analog precoding can beat the hybrid precoding. Simulation results and conclusions are given in Section \uppercase\expandafter{\romannumeral5} and Section \uppercase\expandafter{\romannumeral6}, respectively.
\section{System Model}
We consider a multiuser massive MIMO downlink channel with sub-connected architecture. Multiple users are simultaneously served by a BS which is equipped with $M$ antennas and $K$ RF chains. Each RF chain is connected to a subset of $N$ antennas through dedicated phase shifters where $N=\frac{M}{K}$. Considering that the number of transmit streams should not exceed the number of RF chains, we assume that $K$ single-antenna users are scheduled from the user pool.
Assuming flat Rayleigh fading, the received signal at the $k$th user can be expressed by
\begin{equation}\label{eq:yk}
y_k =\beta_k{\bf h}_k^H{\bf A}{\bf W}{\bf s}+ n_k,\quad k=1,2,\cdots,K,
\end{equation}
where ${\bf h}_k^H \sim {\cal {CN}}({\bf 0}_M,{\bf I}_M)$ denotes the downlink channel from the BS to the $k$th user, and ${\bf s}\in {\mathbb C}^{K\times 1}$ is the data vector with ${\mathbb E}[{\bf s}{\bf s}^H]=\frac{P}{K}{\bf I}_K$ where $P$ is the total transmit power at the BS, $n_k\sim {\cal {CN}}(0,{\sigma ^{\rm{2}}})$ is the additive Gaussian noise, $\beta_k$ denotes the path loss of the $k$th user, and ${\bf A} = [{\bf a}_1,{\bf a}_2,\cdots,{\bf a}_K]\in {\mathbb C}^{M \times K}$ and ${\bf W} = [{\bf w}_1,{\bf w}_2,\cdots,{\bf w}_K]\in{\mathbb C}^{K\times K}$ respectively stand for the analog precoder and the digital precoder. Due to the hardware dissipation caused by the power dividers [14], the analog precoder is written as ${\bf{A}} = \frac{1}{{\sqrt N }}{\bf{F}}$ where ${\bf{F}}$ denotes the equivalent analog precoding implemented by the phase shifter network. To conduct equal power allocation for users, it holds that $\|{\bf F}{\bf w}_k\| = 1$ for $k=1,2,\cdots,K$.

In the current design of hybrid precoding for massive MIMO, the design of analog precoder can be accomplished in two ways, that is, by either utilizing a quantized CSI feedback from users, e.g., precoding matrix indicator (PMI) feedback, or by using uplink channel estimates, e.g., the sounding reference signal (SRS), as specified in the 5G New Radio (NR) specifications [15]. Based on this, the BS designs the analog precoder in terms of discretized phase shifters. Then, an analogly beamformed pilot will be sent by the BS for further equivalent channel estimation to compensate for the performance degradation caused by quantization error of the analog precoder. Each user estimates its own effective channel and feeds back a quantized version of the precoded CSI which is used by the BS to design the digital precoder. In this paper, phases of the $MK$ entries
of $\bf A$ are quantized up to $B_1$ bits, i.e., the codebook is designed as $\mathcal{A}=\left\{ {{e^{\frac{{j2\pi n}}{{{2^{{B_1}}}}}}},n = 0,1, \cdots ,{2^{{B_1}}-1}}\right\}$. Based on the minimum Euclidean distance criterion, a common way of analog precoder design in the
sub-connected structure follows [9] [10]
\begin{equation}\label{eq:analog}
{a_{k,i}} = \left\{ {\begin{array}{*{20}{c}}
{\frac{1}{N}\mathop {{\rm{argmax}}}\limits_{{e^{j{{\hat \varphi }_{k,i}}}} \in\mathcal{A}} {\mathcal{R}}[h_{k,i}^*{e^{j{{\hat \varphi }_{k,i}}}}],}&{N(k - 1) + 1 \le i \le Nk}\\
{0,}&{{\rm{otherwise}}}
\end{array}} \right.
\end{equation}
where $a_{k,i}$ and $h_{k,i}$ are respectively the $i$th element of ${\bf a}_k$ and ${\bf h}_k$, and $\mathcal{R}[x]$ returns the real part of $x$.
For the $k$th user, we define its effective channel as ${\bf g}_k^H \buildrel \Delta \over = {\bf h}_k^H\bf A.$ Then, the ergodic rate of the $k$th user can be written as
\begin{equation}\label{eq:rate}
R_k={\mathbb E}\left[ {{{\log }_2}\left( {1 + \frac{{\frac{{\gamma {\beta _k}}}{K}{{\left| {{\bf{g}}_k^H{{\bf{w}}_k}} \right|}^2}}}{{1 + \frac{\gamma }{K}\sum\limits_{j \ne k} {{\beta _j}{{\left| {{\bf{g}}_k^H{{\bf{w}}_j}} \right|}^2}} }}} \right)} \right],
\end{equation}
where $\gamma\triangleq \frac{P}{\sigma ^{\rm{2}}}$ is the SNR.


The digital part of the system is based on the effective channel. Since the effective channel is always correlated even for independent and identically distributed (i.i.d.) channels of $\bf H$, we adopt a channel statistics-based codebook $\mathcal{G}$ [10]-[12]. The $k$th user quantizes its effective channel ${\bf g}_k$ according to $\hat{{\bf g}}_k={\rm arg~max}_{\hat{\bf g}_{k}^i\in\mathcal{G}}\lvert{\bf g}_k^H\hat{\bf g}_{k}^i\rvert$, in which ${\hat{\bf g}}_{k}^i$ is defined as
\begin{equation}\label{eq:codebook}
{\hat{\bf g}}_{k}^i=\frac{{\bf R}_k^{\frac{1}{2}}{\bf v}_i}{\|{\bf R}_k^{\frac{1}{2}}{\bf v}_i\|},
\end{equation}
where ${\bf R}_k$ denotes the correlation matrix of user $k$'s effective channel and ${\bf v}_i\in {\mathbb C}^{K\times 1}$, $i\in\{1,2,\cdots,2^{B_2}\}$ is an i.i.d. complex Gaussian vector randomly chosen from random vector quantization (RVQ) codebook of size $2^{B_2}$. With the limited feedback of $B_2$ bits, the BS calculates the digital precoder by using the quantized CSI feedback $\hat{\bf g}_k$.
\section{Achievable Rate Analysis}
\subsection{MRT/ZF-based Hybrid Precoding with Limited Feedback}
For the MRT-based hybrid precoding, the digital precoder of user $k$ is determined as
\begin{equation}\label{eq:MRT}
{\bf w}_k={\hat{\bf g}_k}.
\end{equation}
With the design above, we derive the downlink rate in the following theorem.

\emph{Theorem~1:} Tight rate approximation of MRT-based hybrid precoding in the large BS antenna regime is obtained as
\begin{equation}\label{eq:rqmrt}
R_{{\rm H},k}^{{\rm{MRT}}}={\log _2}\left({1 +\frac{{\frac{{\gamma \beta_k }}{K}(\frac{{\pi {\rm{sin}}{{\rm{c}}^2}(\delta )}}{4}+\frac{K}{N}- \frac{{{2^{ - \frac{{{B_2}}}{{K - 1}}}}}}{N})(\frac{{\pi N{\rm{sin}}{{\rm{c}}^2}(\delta )}}{4}+K)}}{{\frac{{\pi N{\rm{sin}}{{\rm{c}}^2}(\delta )}}{4}+K +\frac{{\gamma}}{K}\overline {\beta} _k(\frac{K}{N}+\frac{{\pi {\rm{sin}}{{\rm{c}}^2}(\delta )}}{2})}}}\right),
\end{equation}
where $\delta\buildrel \Delta \over =\frac{\pi}{2^{B_1}}$, sinc$(x)\buildrel \Delta \over =\frac{{\rm sin}(x)}{x}$ and $\overline {\beta} _k  = \sum\limits_{j \ne k} {{\beta _j}}$.
\begin{proof} See Appendix A.\end{proof}
 From the above theorem, we can also get the achievable rate of the MRT-based hybrid system with perfect CSI as a special case by letting $B_1\to\infty$ and $B_2\to\infty$ in (\ref{eq:rqmrt}). It yields
\begin{equation}\label{eq:rmrt}
R_{{\rm H},k}^{\rm MRT_p}={\rm{lo}}{{\rm{g}}_2}\left( {1 + \frac{{\frac{{\gamma {\beta_k}N}}{K}{{(\frac{\pi }{4} + \frac{K}{N})}^2}}}{{\frac{{\pi N}}{4} + K + \frac{{\gamma {{\overline \beta  _k} }}}{K}(\frac{\pi }{2} + \frac{K}{N})}}} \right).
\end{equation}
Note that this result coincides with the result in [9, Eq. (15)] except for that the signal term and the interference term calculated in [9] have an additional factor of $\frac{{\rm{1}}}{N}$ if we let $\beta_i=1, i=1,2,\cdots,K$, which is because we consider the effect of the divider network.

For the ZF-based precoding, the digital precoder equals
\begin{equation}\label{eq:ZF}
{\bf W}^{\rm ZF}=\hat{\bf G}(\hat{\bf G}^H\hat{\bf G})^{-1},
\end{equation}
where $\hat{\bf G}=[\hat{\bf g}_1,\hat{\bf g}_2,\cdots,\hat{\bf g}_K]$ and ${\bf W}^{\rm ZF}=[{\bf w}_1^{\rm ZF},{\bf w}_2^{\rm ZF},\cdots,{\bf w}_K^{\rm ZF}]$. The digital precoder is then normalized as ${\bf w}_k=\frac{{\bf w}_k^{\rm ZF}}{\|{{\bf Fw}_k^{\rm ZF}}\|}$. A tractable and tight bound of its achievable rate is given in the following proposition.

\emph{Proposition~1:} The ergodic rate of the ZF-based hybrid precoding is asymptotically lower bounded by
\begin{equation}\label{eq:rqzf}
R_{{\rm H},k}^{\rm ZF}\geqslant{\rm{lo}}{{\rm{g}}_2}\left( {\frac{{4KN + \pi N\gamma {\beta _k}{\rm{sin}}{{\rm{c}}^2}(\delta )}}{{4KN + 4\gamma {{\overline \beta  }_k}{2^{ - \frac{{{B_2}}}{{K - 1}}}}}}} \right).
\end{equation}
\begin{proof} See Appendix B.\end{proof}
\subsection{Pure Analog Precoding}
When the pure analog precoding is applied, the digital precoding matrix reduces to an identity matrix, i.e., ${\bf W}={\bf I}_K$. Then we can easily get the following proposition.

\emph{Proposition~2:} Tight rate approximation of the pure analog precoding in the large BS antenna regime is characterized as
\begin{equation}\label{eq:ra}
R_{{\rm A},k}={\rm{lo}}{{\rm{g}}_2}\left( {1 + \frac{{\gamma {\beta _k}(\frac{{\pi N{\rm{sin}}{{\rm{c}}^2}(\delta )}}{4} + 1)}}{{KN + \gamma {{\overline \beta  }_k}}}} \right).
\end{equation}
\begin{proof} With the pure analog precoding, ${\mathbb E}[{\left| {{{\bf{g}}_k^H}{{\bf{w}}_k}} \right|^2}]={\mathbb E}[{\left| {{g_{k,k}}} \right|^2}]=\frac{{\pi {\rm{sin}}{{\rm{c}}^2}{\rm{(}}\delta {\rm{)}}}}{4} + \frac{{{1} }}{N}-\frac{{\pi {\rm{sin}}{{\rm{c}}^2}{\rm{(}}\delta {\rm{)}}}}{4N}$ and ${\mathbb E}[{\left| {{{\bf{g}}_k^H}{{\bf{w}}_j}} \right|^2}]={\mathbb E}[{\left| {{g_{k,i}}} \right|^2}]=\frac{1}{N}$. By applying [16, Lemma 1], and substituting these two expressions into (\ref{eq:rate}) and using similar manipulations as that in the last step of (\ref{eq:signal}), the proof completes.\end{proof}
As a special case, we have the achievable rate of the system with perfect CSI by letting $B_1\to\infty$, which gives
\begin{equation}\label{eq:ra1}
R_{{\rm A},k}^{\rm p}={\rm{lo}}{{\rm{g}}_2}\left( {1 + \frac{{\frac{{\pi N\gamma {\beta _k}}}{4}}}{{KN + \gamma {{\overline \beta  }_k}}}} \right),
\end{equation}
which consists with the existing result in [9, Eq. (13)].
\section{Pure Analog vs Hybrid Precoding}
To compare the performance of the pure analog precoding and the hybrid precoding, we evaluate the rate gap as
\begin{equation}\label{eq:gap}
\Delta R=R_{{\rm H},k}^{X}-R_{{\rm A},k},
\end{equation}
where ${{X}} \in \{ {\rm{ZF}},{\rm{MRT}}\}$ represents the strategy of the digital precoding in the hybrid precoding.
\vspace{-10pt}
\subsection{MRT-based Hybrid Precoding}
By letting $\triangle R \geq 0$, and substituting (\ref{eq:rqmrt}) and (\ref{eq:ra}) into (\ref{eq:gap}), we get
\begin{equation}\label{eq:condition1}
\frac{{\gamma{\overline \beta  _k}}}{K}\left( {\frac{{K - 2 - {2^{ - \frac{{{B_2}}}{{K - 1}}}}}}{N} - \frac{{\pi {\rm{sin}}{{\rm{c}}^2}(\delta )}}{4}} \right) > {2^{ - \frac{{{B_2}}}{{K - 1}}}} - K + 1.
\end{equation}
It implies that the MRT-based hybrid precoding outperforms the pure analog precoding if (\ref{eq:condition1}) holds. It is obvious that the right hand side of (\ref{eq:condition1}) is negative when $K>1$. Checking that $\frac{{K - 2 - {2^{ - \frac{{{B_2}}}{{K - 1}}}}}}{{{N}}}\geq\frac{{\pi {\rm{sin}}{{\rm{c}}^2}{\rm{(}}\delta {\rm{)}}}}{{4}}$, we can rewrite (\ref{eq:condition1}) as
\begin{equation}\label{eq:condition11}
K \geq \frac{{\pi N{\rm{sin}}{{\rm{c}}^2}(\delta )}}{4}{\rm{ + }}{2^{ - \frac{{{B_2}}}{{K - 1}}}} + 2,
\end{equation}
under which the MRT-bsed hybrid precoding outperforms the pure analog precoding for all SNR values. This is due to the fact that multiuser interference becomes a dominant factor when $K$ becomes large, and the hybrid precoding which utilizes digital processing to eliminate interference enjoys a better performance than the pure analog precoding.

Moreover from (\ref{eq:condition11}), it is revealed that $B_2$ has a marginal impact on the performance of the MRT-based hybrid precoding since ${{2^{ - \frac{{{B_2}}}{{K - 1}}}}} \ll1$. Therefore we can focus on the impact of $B_1$ on system performance.
By applying Taylor's expansion to ${\sin ^2}(\delta )$ with $\delta  = \frac{\pi }{{{2^{{B_1}}}}}$, we get ${\rm{sin}}{{\rm{c}}^2}(\frac{\pi }{{{2^{{B_1}}}}}) \approx 1 - \frac{{{{\left( {\frac{\pi }{{{2^{{B_1}}}}}} \right)}^2}}}{3}$, and (\ref{eq:condition11}) is further equivalent to
\begin{equation}\label{eq:condition12}
{B_1} \leq \frac{1}{2}{\log _2}\left( {\frac{{{\pi ^3}N}}{{3\pi N - 12\left( {K - 2 - {2^{ - \frac{{{B_2}}}{{K - 1}}}}} \right)}}} \right) \buildrel \Delta \over = B_1^0.
\end{equation}
Otherwise, when $B_1>B_1^0$, it is not difficult to get
\begin{equation}\label{eq:condition_mrt}
\Delta R\left\{ {\begin{array}{*{20}{c}}
{ > 0,}&{\gamma  < {\gamma _{0}}}\\
{ \le 0,}&{\gamma  \geq {\gamma _{0}}}
\end{array}} \right.
\end{equation}
where $\gamma_0\buildrel \Delta \over =\frac{{M\left( {{2^{ - \frac{{{B_2}}}{{K - 1}}}} - K + 1} \right)}}{{{\overline \beta  _k}\left( {K - 2 - {2^{ - \frac{{{B_2}}}{{K - 1}}}} - \frac{{\pi N{\rm{sin}}{{\rm{c}}^2}(\delta )}}{4}} \right)}}$.

\emph{Remark~1:} When $B_1\leq B_1^0$, the MRT-bsed hybrid precoding always outperforms the pure analog precoding in terms of the ergodic achievable rate for all SNRs. Otherwise the MRT-bsed hybrid precoding can beat the pure analog precoding only for low SNRs satisfying $\gamma<\gamma _0$.
\subsection{ZF-based Hybrid Precoding}
Similarly by letting $\triangle R \geq 0$ for $X=\rm ZF$ in (\ref{eq:gap}) and using (\ref{eq:rqzf}) and (\ref{eq:ra}), we obtain
\begin{equation}\label{eq:condition21}
{B_2} > (K - 1){\log _2}\left( {1 + \frac{{4({\beta _k} + {{\overline \beta  }_k})}}{{\pi N{\beta _k}{\rm{sin}}{{\rm{c}}^2}(\delta )}}} \right)\buildrel \Delta \over = B_{\rm{2}}^{\rm{0}},
\end{equation}which uses the fact that $\frac{{\pi N{\beta _k}{\rm{sin}}{{\rm{c}}^2}(\delta )\left( {{\rm{1 - }}{2^{ - \frac{{{B_2}}}{{K - 1}}}}} \right)}}{4} > {2^{ - \frac{{{B_2}}}{{K - 1}}}}({\beta _k} + {\overline \beta  _k})$. When $B_2>B_2^0$, it holds true that
\begin{equation}\label{eq:condition_zf}
\Delta R\left\{ {\begin{array}{*{20}{c}}
{ > 0,}&{\gamma> {\gamma _1}}\\
{ \le 0,}&{\gamma  \leq {\gamma _1}}
\end{array}} \right.
\end{equation}
where $\gamma_1\buildrel \Delta \over =\frac{{4M({{\overline \beta  }_k}{2^{ - \frac{{{B_2}}}{{K - 1}}}} + {\beta _k})}}{{{{\overline \beta  }_k}\left( {\pi N{\beta _k}{\rm{sin}}{{\rm{c}}^2}(\delta )\left( {{\rm{1 - }}{2^{ - \frac{{{B_2}}}{{K - 1}}}}} \right) - 4({\beta _k} + {{\overline \beta  }_k}){2^{ - \frac{{{B_2}}}{{K - 1}}}}} \right)}}$.

Otherwise, when $B_2\leq B_2^0$, the pure analog precoding always outperforms the ZF-based hybrid precoding for all SNRs. This is because the precoding vector of the $k$th user is supposed to lie in the null space of channels of other users and thus the ZF-based hybrid precoding is more dependent on the accuracy of channel estimation to cancel out interference.

\emph{Remark~2:} If $B_2\leq B_2^0$, the pure analog precoding beats the ZF-based hybrid precoding for all SNRs. Otherwise the ZF-based hybrid precoding takes over for high SNRs satisfying $\gamma\geq\gamma_1$.
\section{Simulation Results}
In this section, we compare the downlink performance of the analog precoding and hybrid precoding based on different linear processing schemes. The path loss factor $\beta_k,~k=1,2,\cdots,K$ is chosen uniformly in [0.5, 1.5].
\begin{figure}[!t]
\setlength{\abovecaptionskip}{0cm}
\setlength{\belowcaptionskip}{-0.5cm}
\centering
\subfigure[$M=120$, $K=6$, and $B_2=10.$]{
\begin{minipage}[t]{0.5\linewidth}
\centering
\includegraphics[width=3.9in]{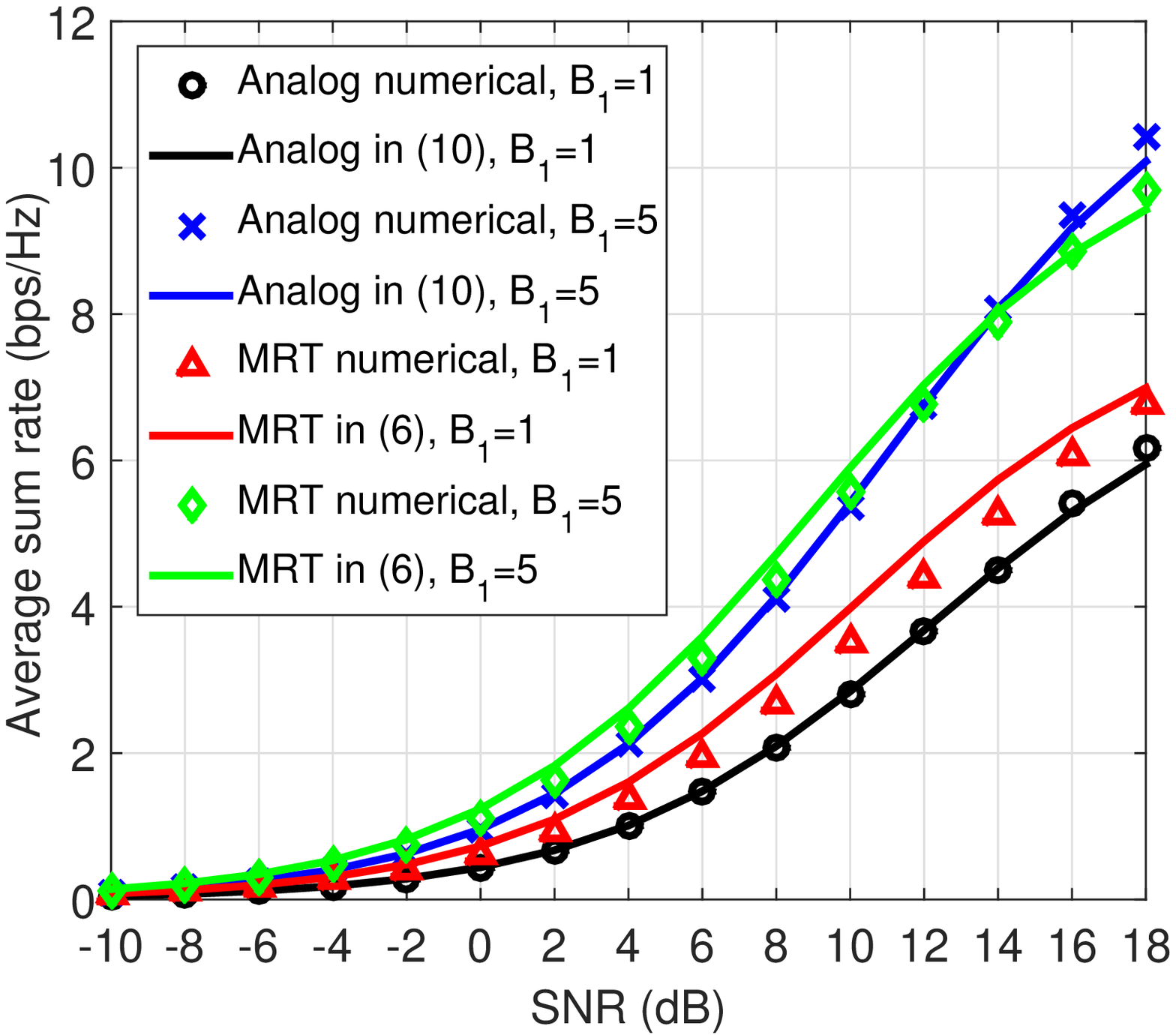}
\end{minipage}%
}%
\subfigure[$M=60$, $K=6$, and $B_1=2$.]{
\begin{minipage}[t]{0.5\linewidth}
\centering
\includegraphics[width=3.9in]{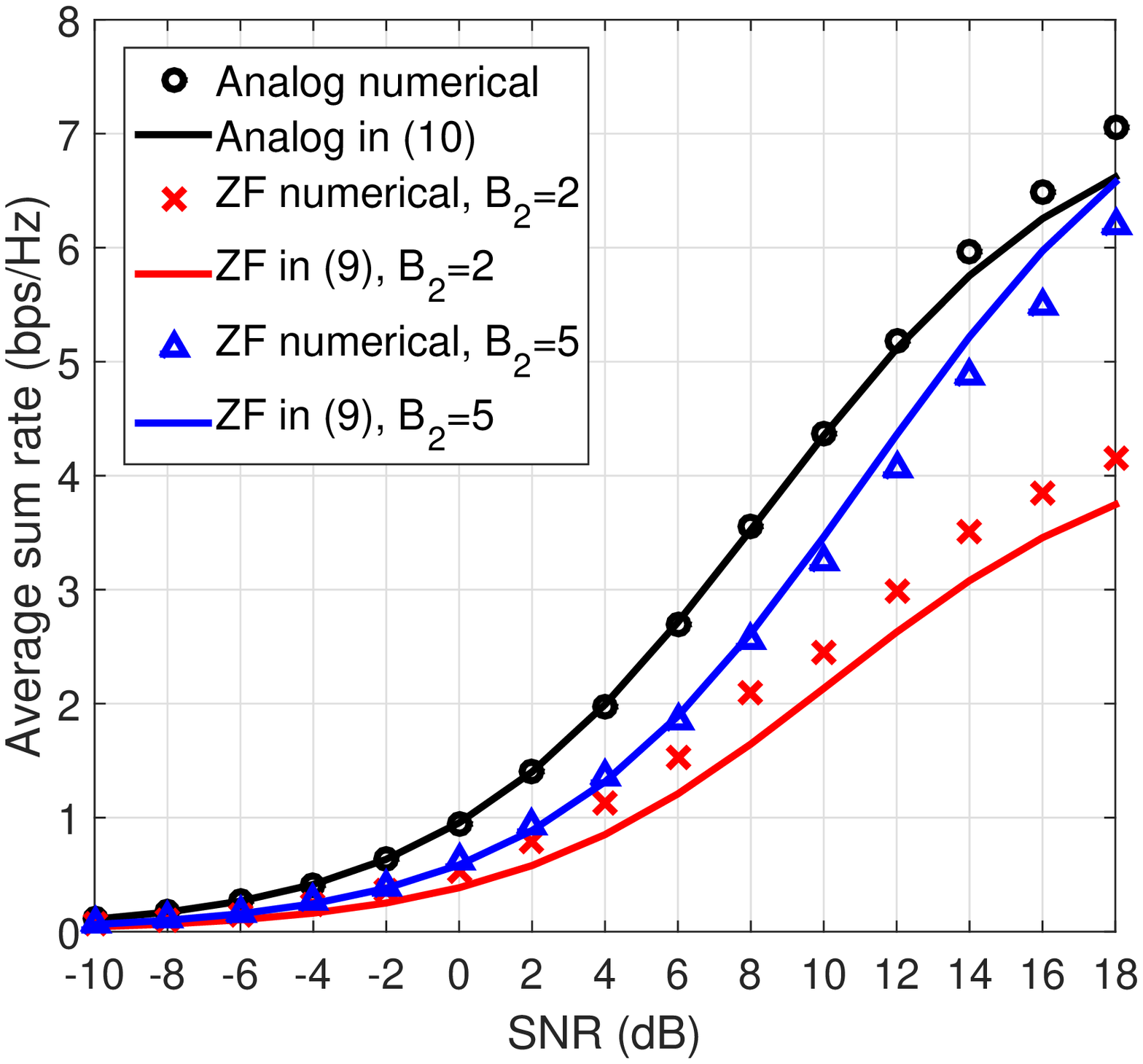}
\end{minipage}%
}%
\centering
\caption{Achievable rates of analog precoding and MRT-based hybrid precoding in (a) and ZF-based hybrid precoding in (b)}
\end{figure}
\subsection{Rayleigh Fading Channels}

In Fig. 1(a) we illustrate the spectral efficiency achieved by pure analog precoding and the MRT-based hybrid precoding. It is observed that when $B_1\leq B_1^0=1.05$, the MRT-based hybrid precoding outperforms pure analog precoding for all SNRs, while when $B_1> B_1^0$ the pure analog precoding performs better at high SNRs, confirming the conclusions in Section \uppercase\expandafter{\romannumeral4}. This is because for small $B_1$, the system performance is severely affected by the quantization error, i.e., the interference becomes dominant, thus the hybrid precoding using digital processing to eliminate interference outperforms the pure analog precoding.  In addition, the corresponding threshold calculated from (14) is 17 for $B_1=5$, so for $K=6<17$, the pure analog precoding outperforms the MRT-based hybrid precoding at high SNRs.

Similar trends can be found when comparing pure analog precoding and the ZF-based hybrid precoding in Fig. 1(b). When the effective
channel is roughly quantized, i.e., $B_2\leq B_2^0=4.29$, the pure analog precoding performs better for all SNRs since the accuracy of channel estimation is insufficient to support effective beamforming for ZF, while the ZF-based hybrid precoding takes over at high SNRs when $B_2> B_2^0$, which verifies our observations in Section \uppercase\expandafter{\romannumeral4}.
\subsection{Large mmWave Multiuser Channels}
Apart from Rayleigh fading channels, hybrid/analog precoding can also be applied to mmWave communications. To capture the nature of high-frequency propagations, we adopt the widely-used geometric channel model [7] [9]
\begin{equation}\label{eq:mmwave}
{\bf{h}}_k^{\rm{H}} = \sqrt {\frac{{{M}}}{{{N_p}}}} \sum\limits_{l = 1}^{{N_p}} {\alpha _l^k{{\bf{a}}^H}} (\phi _l^k),
\end{equation}
where $N_p$ is the number of propagation paths from BS to user and ${\alpha _l^k}\sim {\cal {CN}}(0,1)$ denotes the complex gain of the $l$-th path. $\phi _l^k$ is the azimuth angle of departure drawn independently from the uniform distribution over [$0, 2\pi$]. ${\bf{a}}(\phi _l^k)$ is the array response vector of BS. Here we consider a uniform
linear array (ULA) whose array response vector is given by [9, eq. (29)].

Fig. 2 shows that conclusions on the superiority of the pure analog precoding over hybrid precodings also hold for the mm-Wave channels. Specifically, when $B_2$ is small, i.e., $B_2=3$, the analog precoding outperforms the ZF-based hybrid precoding for all SNRs, while for large $B_2$, i.e., $B_2=12$, the ZF-based hybrid precoding takes over at high SNRs, which verifies Remark 2.
\begin{figure}[!t]
\setlength{\abovecaptionskip}{0cm}
\setlength{\belowcaptionskip}{-0.5cm}
\centering 
\includegraphics[width=6.0in]{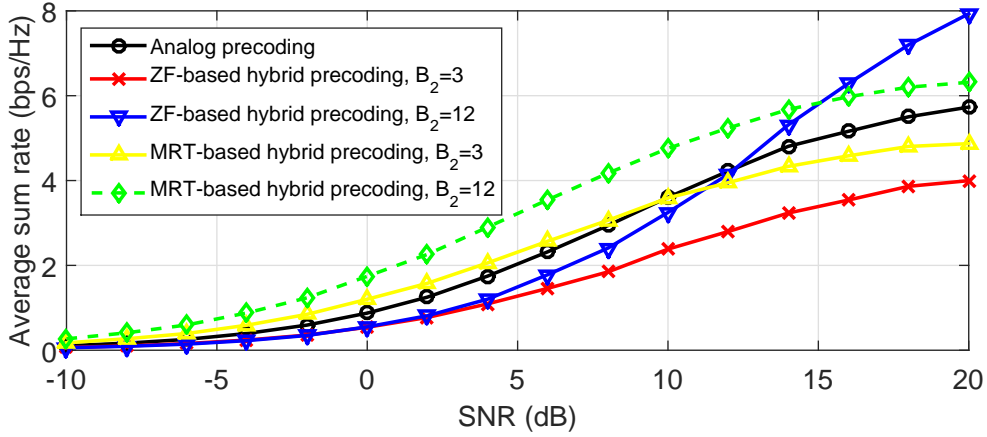} 
\caption{Achievable rates over mmWave channels with $M=40, K=5, B_1=2$, and $N_p=10$.}
\label{fig:0}       
\end{figure}
\section{Conclusion}
In this paper, we derive tight rate approximations of a massive MIMO system using pure analog precoding and hybrid precoding with limited feedback in the large BS antenna regime. The effect of quantized analog and digital precodings are characterized in the obtained expressions. Furthermore, it is revealed that pure analog precoding outperforms hybrid precoding in terms of the ergodic achievable rate under certain conditions, which have been derived in closed forms with respect to the SNR, the number of users and the number of feedback bits. Numerical results verify the observed.
\begin{appendices}
\section{Proof Of Theorem 1}
To evaluate (\ref{eq:rate}), we calculate the terms ${\mathbb E}\left[{\left| {{{\bf{g}}_k^H}{{\bf{w}}_k}} \right|^2}\right]$ and ${\mathbb E}\left[{\left| {{{\bf{g}}_k^H}{{\bf{w}}_j}} \right|^2}\right]$. Let ${{\bf{g}}_k} = \left\| {{{\bf{g}}_k}} \right\|{\tilde {\bf{g}}_k}$ where ${\tilde {\bf{g}}_k}$ is the normalized ${\bf{g}}_k$, then we have ${\mathbb E}\left[{\left| {{{\bf{g}}_k^H}{{\bf{w}}_k}} \right|^2}\right]={\mathbb E}\left[\left\| {{{\bf{g}}_k}} \right\|_{\rm F}^2\right]{\mathbb E}\left[{\left| {{\tilde{\bf{g}}_k^H}{{\bf{w}}_k}} \right|^2}\right]$.

From the definition of ${\bf{g}}_k$, its $k$th element equals
\begin{equation}\label{eq:gkk}
{g_{k,k}} = \frac{1}{{N}}\sum\limits_{l = N(k - 1) + 1}^{Nk} {{h_{k,l}}{e^{ - j{{\hat \varphi }_{k,l}}}}}\overset{(\rm a)}=\frac{1}{{N }}\sum\limits_{l = N(k - 1) + 1}^{Nk} {\lambda _l},
\end{equation}where (a) uses ${\lambda _l} \buildrel \Delta \over ={{h_{k,l}}{e^{ - j{{\hat \varphi }_{k,l}}}}}$. Defining ${\varepsilon _{k,l}} \buildrel \Delta \over = {\varphi _{k,l}} - {\hat \varphi _{k,l}}$, it yields ${\lambda _l} ={|{h_{k,l}}|{e^{j{{\varepsilon }_{k,l}}}}}$. Recall ${\bf h}_k^H \sim {\cal {CN}}({\bf 0}_M,{\bf I}_M)$, where ${\rm{\{ }}{{h}_{k,i}}\} $'s are i.i.d. complex Gaussian variables with zero mean and unit variance. It implies that $| {{h_{k,l}}}|$ follows the Rayleigh distribution with mean $\frac{{\sqrt \pi  }}{2}$ and variance $1 - \frac{\pi }{4}$. The quantization error ${\varepsilon _{k,l}}\sim U[-\delta,\delta)$ is a uniform distribution with $\delta\buildrel \Delta \over =\frac{\pi }{{{2^{{B_1}}}}}$. Then we have
\begin{equation}\label{eq:gkk_rm}
{\mathbb E}[\mathcal{R}[{\lambda _l}]]={\mathbb E}\left[\mathcal{R}[{|{h_{k,l}}|{e^{j{{\varepsilon }_{k,l}}}}}]\right]={\mathbb E}[| {{h_{k,l}}}|\cos ({\varepsilon _{k,l}})]
\overset{(\rm a)}={\mathbb E}[| {{h_{k,l}}}|]{\mathbb E}[\cos ({\varepsilon _{k,l}})]\overset{(\rm b)}=\frac{{\sqrt \pi  {\rm{sinc}}(\delta )}}{2},
\end{equation}
where (a) utilizes the independence between $| {{h_{k,l}}}|$ and ${\varepsilon _{k,l}}$, and (b) is due to ${\mathbb E}[| {{h_{k,l}}}|]=\frac{{\sqrt \pi  }}{2}$ and ${\mathbb E}[\cos ({\varepsilon _{k,l}})]=\frac{1}{{2\delta }}\int_{ - \delta }^\delta  {\cos {\varepsilon _{k,l}}} d{\varepsilon _{k,l}}={\rm{sinc}}(\delta )$ because of the given distributions of $| {{h_{k,l}}}|$ and ${\varepsilon _{k,l}}$. Analogously, we can also get ${\mathbb E}[(\mathcal{R}[{\lambda _l}])^2]={\mathbb E}[| {{h_{k,l}}}|^2]{\mathbb E}[\cos^2 ({\varepsilon _{k,l}})]=\frac{1+{\rm{sinc}}(\delta )\cos (\delta )}{2}$. Then,
\begin{equation}\label{eq:gkk_rv}
{\mathbb V}[\mathcal{R}[{\lambda _l}]]={\mathbb E}[(\mathcal{R}[{\lambda _l}])^2]-({\mathbb E}[\mathcal{R}[{\lambda _l}]])^2=w_1,
\end{equation}
where ${w_1}\buildrel \Delta \over =\frac{1+{\rm{sinc}}(\delta )\cos (\delta )}{2}-\frac{{\pi {\rm{sin}}{{\rm{c}}^2}(\delta )}}{4}$. Applying the Central Limit Theorem to (\ref{eq:gkk}) and using (\ref{eq:gkk_rm}) and (\ref{eq:gkk_rv}), we get
\begin{equation}\label{eq:gkk_rd}
\mathcal{R}[g_{k,k}] \sim \mathcal{N}\left(\frac{{\sqrt \pi  {\rm{sinc(}}\delta {\rm{)}}}}{2},\frac{{{w_1}}}{N}\right).
\end{equation}
Following similar reasons, it can be readily proved that
\begin{equation}\label{eq:gkk_id}
\mathcal{I}[g_{k,k}] \sim \mathcal{N}(0,\frac{{{w_2}}}{N}),~\!
\mathcal{R}[g_{k,i}] \sim \mathcal{N}(0,\frac{{{1}}}{2N}),~\!\mathcal{I}[g_{k,i}] \sim \mathcal{N}(0,\frac{{{1}}}{2N}),
\end{equation}
where ${w_2}\buildrel \Delta \over = \frac{1-{\rm{sinc}}(\delta )\cos (\delta )}{2}$ and $\mathcal{I}[x]$ returns the imaginary part of $x$. Now we have
\begin{align}
\label{eq:gkk_val}{\mathbb E}[{\left| {{g_{k,k}}} \right|^2}]&={\mathbb E}[{\left| {{\cal R}[{g_{k,k}}]} \right|^2}]\!+\!{\mathbb E}[{\left| {{\cal I}[{g_{k,k}}]} \right|^2}]=\frac{{\pi {\rm{sin}}{{\rm{c}}^2}{\rm{(}}\delta {\rm{)}}}}{4}\!+\!\frac{{{w_1}\!+\!{w_2}}}{N}.\\
\label{eq:gki_val}{\mathbb E}[{\left| {{g_{k,i}}} \right|^2}]&={\mathbb E}[{\left| {{\cal R}[{g_{k,i}}]} \right|^2}]+{\mathbb E}[{\left| {{\cal I}[{g_{k,i}}]} \right|^2}]=\frac{1}{N}.
\end{align}
\indent According (\ref{eq:gkk_val}) and (\ref{eq:gki_val}), it gives
\begin{equation}\label{eq:gks}
{\mathbb E}\left[\left\| {{{\bf{g}}_k}} \right\|_{\rm F}^2\right]\!=\!{\mathbb E}[{\left| {{g_{k,k}}} \right|^2}]+\!\!\!\!\!\!{\sum\limits_{i = 1,i \ne k}^K\!\!\!\!{{\mathbb E}[\left| {{g_{k,i}}} \right|} ^2}]\!=\!\frac{{\pi {\rm{sin}}{{\rm{c}}^2}{\rm{(}}\delta {\rm{)}}}}{4}+\!\frac{K}{N}-\!\frac{{\pi {\rm{sin}}{{\rm{c}}^2}{\rm{(}}\delta {\rm{)}}}}{{4N}}.
\end{equation}
From [10, Lemma 2], ${\bf R}_k$ can be asymptotically written as ${\bf R}_k={\rm diag}{[r_{1,1},r_{2,2},\cdots,r_{k,k}]}$ where
\begin{equation}\label{eq:Rk}
\begin{split}
r_{i,i}=\left\{\begin{array}{lc}
{\frac{{\pi {\rm{sin}}{{\rm{c}}^2}(\delta )}}{4} + \frac{1}{N} - \frac{{\pi {\rm{sin}}{{\rm{c}}^2}(\delta )}}{{4N}}},\quad &i=k\\
\frac{1}{N},\quad &i\neq k\end{array}\right.
\end{split}
\end{equation}
and diag[$\cdot$] returns a diagonal matrix with the input as its elements. Then we follow the result in [11] and get
\begin{equation}\label{eq:gkss}
{\mathbb E}\left[{\left| {{\tilde{\bf{g}}_k^H}{\hat{\bf{g}}_k}} \right|^2}\right]  \approx  1-\frac{{\sigma _{k,2}^2}}{{\sigma _{k,1}^2}}{2^{-\frac{B_2}{K-1}}}=1 - \frac{{{2^{ - \frac{{{B_2}}}{{K - 1}}}}}}{{\frac{{\pi N{\rm{sin}}{{\rm{c}}^2}(\delta )}}{4} + 1 - \frac{{\pi {\rm{sin}}{{\rm{c}}^2}(\delta )}}{4}}},
\end{equation}
where ${\sigma _{k,1}}$ and ${\sigma _{k,2}}$ respectively stand for the largest and the second largest singular value of ${\bf R}^{\frac{1}{2}}_k$.

Combining (\ref{eq:gks}) and (\ref{eq:gkss}), it yields
\begin{equation}\label{eq:signal}
{\mathbb E}[{\left| {{{\bf{g}}_k^H}{{\bf{w}}_k}} \right|^2}]\overset{(\rm a)}={\mathbb E}[\left\| {{{\bf{g}}_k}} \right\|_F^2]{\mathbb E}[{\left| {{\tilde{\bf{g}}_k^H}{\hat{\bf{g}}_k}} \right|^2}]\nonumber\\
\overset{(\rm b)}{\to}\frac{{\pi {\rm{sin}}{{\rm{c}}^2}(\delta )}}{4} + \frac{K}{N} - \frac{{{2^{ - \frac{{{B_2}}}{{K - 1}}}}}}{N},
\end{equation}
where (a) utilizes ${{\bf{w}}_k} = {{\bf{g}}_k}$ and in (b) we consider the fact that $\frac{{\frac{{\pi {\rm{sin}}{{\rm{c}}^2}(\delta )}}{4} + \frac{K}{N}}}{{\frac{{\pi {\rm{sin}}{{\rm{c}}^2}(\delta )}}{4} + \frac{K}{N} - \frac{{\pi {\rm{sin}}{{\rm{c}}^2}(\delta )}}{{4N}}}} \to 1$ and $\frac{{\frac{{\pi {\rm{sin}}{{\rm{c}}^2}(\delta )}}{4} + \frac{K}{N} - \frac{{\pi {\rm{sin}}{{\rm{c}}^2}(\delta )}}{{4N}}}}{{\frac{{\pi N{\rm{sin}}{{\rm{c}}^2}(\delta )}}{4} + 1 - \frac{{\pi {\rm{sin}}{{\rm{c}}^2}(\delta )}}{4}}} \to \frac{1}{N}$ hold in massive MIMO with large $N$ and fixed $K$.

From (\ref{eq:codebook}), we get ${\hat {g}_{j,j}} = \frac{{{{({\bf{R}}_j^{\frac{1}{2}})}_{j,j}}{{{v}}_{i,j}}}}{\|{{\bf{R}}_j^{\frac{1}{2}}{{\bf{v}}_{i}}}\|}$, where ${\hat {g}_{j,j}}$ and ${v_{i,j}}$ are respectively the $j$th element of ${\hat {\bf{g}}_j}$ and ${{\bf{v}}_i}$, and $i$ denotes the index of the quantization vector of ${\bf{g}}_j$. Then we have
\begin{equation}\label{eq:g^jj_m}
{\mathbb E}[{{\hat g}}_{j,j}]= {\mathbb E}\left[\frac{{{{({\bf{R}}_j^{\frac{1}{2}})}_{j,j}}{{{v}}_{i,j}}}}{\|{{\bf{R}}_j^{\frac{1}{2}}{{\bf{v}}_{i}}}\|}\right]\overset{(\rm a)}={\mathbb E}[{{v_{i,j}}}]{\mathbb E}\left[\frac{{{{({\bf{R}}_j^{\frac{1}{2}})}_{j,j}}}}{{\| {{\bf{R}}_j^{\frac{1}{2}}{{\bf{v}}_i}}\|}}\right]\overset{(\rm b)}=0,
\end{equation}
where (a) uses the independence between ${v_{i,j}}$ and $\frac{{{{({\bf{R}}_j^{\frac{1}{2}})}_{j,j}}}}{{\| {{\bf{R}}_j^{\frac{1}{2}}{{\bf{v}}_i}}\|}}$, and (b) results from ${\bf v}_{i}\sim {\cal {CN}}({\bf 0}_M,{\bf I}_M)$. Analogously,
\begin{equation}\label{eq:g^jj_2}
{\mathbb E}[|{{\hat g}}_{j,j}|^2]= {\mathbb E}\left[\frac{{{{({\bf{R}}_j)}_{j,j}}{|{{v_{i,j}}}|^2}}}{\|{{\bf{R}}_j^{\frac{1}{2}}{{\bf{v}}_{i}}}\|^2}\right]={\mathbb E}[|{{v_{i,j}}}|^2]\frac{{{{({\bf{R}}_j)}_{j,j}}}}{{\mathbb E}\left[{\| {{\bf{R}}_j^{\frac{1}{2}}{{\bf{v}}_i}}\|^2}\right]}\overset{(\rm a)}=\frac{{\frac{{\pi {\rm{sin}}{{\rm{c}}^2}(\delta )}}{4} + \frac{1}{N} - \frac{{\pi {\rm{sin}}{{\rm{c}}^2}(\delta )}}{{4N}}}}{{\frac{{\pi {\rm{sin}}{{\rm{c}}^2}(\delta )}}{4} + \frac{K}{N} - \frac{{\pi {\rm{sin}}{{\rm{c}}^2}(\delta )}}{{4N}}}},
\end{equation}
where (a) uses ${\mathbb E}\left[{\| {{\bf{R}}_j^{\frac{1}{2}}{{\bf{v}}_i}} \|^2}\right]=\sum\limits_{l = 1}^K {{{({{\bf{R}}_j})}_{l,l}}{\mathbb E}[|{v_{i,l}}|^2]}=\sum\limits_{l = 1}^K {{{({{\bf{R}}_j})}_{l,l}}} ={\frac{{\pi {\rm{sin}}{{\rm{c}}^2}(\delta )}}{4} + \frac{K}{N} - \frac{{\pi {\rm{sin}}{{\rm{c}}^2}(\delta )}}{{4N}}}$ and ${\mathbb E}[|{{v_{i,l}}}|^2]=1$ according to the distribution of ${\bf v}_i$ and (\ref{eq:Rk}). Following trivially the above steps, we get ${\mathbb E}[{{\hat g}}_{j,i}]=0$, and ${\mathbb E}[|{{\hat g}}_{j,i}|^2]=\frac{1}{{\frac{{\pi N{\rm{sin}}{{\rm{c}}^2}(\delta )}}{4} + K - \frac{{\pi {\rm{sin}}{{\rm{c}}^2}(\delta )}}{4}}}$.
Providing that ${\rm{\{ }}{{v}_{i,l}}\} $'s are i.i.d. complex Gaussian variables with zero mean and unit variance, then ${\hat {g}_{j,i}}$ and ${\hat {g}_{j,l}}$ are independent for any $i\neq l$, which implies ${\mathbb E}[{{\hat g}}_{j,i}{{\hat g}^*_{j,l}}]={\mathbb E}[{{\hat g}}_{j,i}]{\mathbb E}[{{\hat g}^*}_{j,l}]=0$ and ${\mathbb E}[{\left| {{{\hat g}}_{j,i}} \right|^2}{| {{{\hat g}^*_{j,l}}}|^2}]={\mathbb E}[{\left| {{{\hat g}_{j,i}}} \right|^2}]{\mathbb E}[{|{{{\hat g}^*_{j,l}}} |^2}]$. Then ${\mathbb E}\left[{\left| {{{\bf{g}}_k^H}{{\bf{w}}_j}} \right|^2}\right]$ can be calculated as
\begin{align}\label{eq:intf}
{\mathbb E}\left[{\left| {{{\bf{g}}_k^H}{{\bf{w}}_j}}\right|^2}\right]&={\mathbb E}\left[{\left| {{{\bf{g}}_k^H}{\hat{\bf{g}}_j}}\right|^2}\right]={\mathbb E}\left[{\left|\sum\limits_{i = 1}^K {{g^*_{k,i}}} {\hat g_{j,i}}\right|^2}\right]\overset{(\rm a)}=\sum\limits_{1 \le i \le l \le K}^K {{\mathbb E}[g_{k,i}^*{g_{k,l}}]{\mathbb E}[{{\hat g}_{j,i}}\hat g_{j,l}^*]}\nonumber\\
&\overset{(\rm b)}=\sum\limits_{i = 1}^K{\mathbb E}[{\left| {{{\hat g}_{j,i}}} \right|^2}]{\mathbb E}[{\left| {{g_{k,i}}} \right|^2}]\overset{(c)}{\to}\frac{{\frac{{\pi {\rm{sin}}{{\rm{c}}^2}(\delta )}}{2} + \frac{K}{N}}}{{\frac{{\pi N{\rm{sin}}{{\rm{c}}^2}(\delta )}}{4} + K}},
\end{align}
where (a) is due to the independence between ${\bf{g}}_k$ and ${\hat{\bf{g}}_j}$, (b) holds because ${\mathbb E}[g_{k,i}^*{g_{k,l}}]{\mathbb E}[{{\hat g}}_{j,i}{{\hat g}^*_{j,l}}]=0$ for any $i\neq l$, and (c) is obtained by using similar manipulations as that in the last step of (\ref{eq:signal}). Finally, by applying [16, Lemma 1] and substituting (\ref{eq:signal}) and (\ref{eq:intf}) into (\ref{eq:rate}), the theorem is proved.
\section{Proof Of Proposition 1}
From the distributions of $g_{k,k}$ and $g_{k,i}$ in (\ref{eq:gkk_rd}) -- (\ref{eq:gkk_id}), we get $g_{k,k}\overset{a.s.}{\to}\frac{{\sqrt \pi  {\rm{sinc}}(\delta )}}{2}$ and $g_{k,i}\overset{a.s.}{\to}0$ since $\frac{{{w_1}}}{N} \to 0,\frac{{{w_2}}}{N} \to 0,$ and $\frac{1}{{2N}} \to 0$ for the massive MIMO with large $N$, thus ${\bf G}\overset{a.s.}{\to}\frac{{\pi {\rm{sin}}{{\rm{c}}^2}(\delta )}}{4}{{\bf{I}}_K}$. Now the asymptotic rate of the ZF-based hybrid system with perfect CSI can be represented as
\begin{equation}\label{eq:ZF}
R_{{\rm H},k}^{\rm ZF_p}\overset{a.s.}{\to}{\rm{lo}}{{\rm{g}}_2}\left( {1 + \frac{{\gamma\beta_k \pi {\rm{sin}}{{\rm{c}}^2}(\delta )}}{{4K}}} \right).
\end{equation}
On the other hand, it is known from [12] that the rate loss caused by the imperfect CSI can be upper bounded by ${\rm{lo}}{{\rm{g}}_2}\left( {1 + \frac{{\gamma{\overline \beta _k}{\mathbb E}\left[{{\left| {{\bf{g}}_k^H{{\bf{w}}_j}} \right|}^2}\right]}}{K}} \right)$. Thanks to the orthogonality between $\hat{\bf g}_k$ and ${\bf w}_j$ in the ZF precoding, we further have
\begin{equation}\label{eq:orth}
{\mathbb E}\left[{\left| {{\bf{g}}_k^{\rm{H}}{{\bf{w}}_j}} \right|^2}\right] \le {\mathbb E}\left[\left\| {{{\bf{g}}_k}} \right\|_{\rm F}^2\right]{\mathbb E}\left[1 - {\left| {\tilde {\bf{g}}_k^H{{\hat {\bf{g}}}_k}} \right|^2}\right] = \frac{{{2^{ - \frac{{{B_2}}}{{K - 1}}}}}}{N}.
\end{equation}
Therefore, we have
\begin{equation}\label{eq:rloss}
R_{\rm loss}\leq{\log _2}(1 + \frac{{\gamma{\overline \beta  _k} {2^{ - \frac{{{B_2}}}{{K - 1}}}}}}{{KN}}).
\end{equation}
Combining (\ref{eq:ZF}) and (\ref{eq:rloss}), the proposition is directly proved.
\end{appendices}

\end{document}